\documentclass[aps,pra,amsmath,showpacs,twocolumn,floatfix,superscriptaddress]{revtex4}
\usepackage{bm}
\usepackage{graphicx}
\begin{document}
\title{Second-order superposition operations via Hong-Ou-Mandel interference}

\author{Su-Yong Lee}
\affiliation{Department of Physics, Texas A\&M University at Qatar,
 POBox 23874, Doha, Qatar}

\author{Hyunchul Nha}
\affiliation{Department of Physics, Texas A\&M University at Qatar,
 POBox 23874, Doha, Qatar}
\affiliation{Institute f{\"u}r Quantenphysik, Universit{\"a}t Ulm, D-89069 Ulm, Germany}
\date{\today}

\begin{abstract}
We propose an experimental scheme to implement a second-order nonlocal superposition operation $\hat{a}^{\dag 2}+e^{i\phi}\hat{b}^{\dag 2}$ and its variants by way of Hong-Ou-Mandel interference.  The second-order coherent operations enable us to generate a NOON state with high particle number in a heralded fashion and also can be used to enhance the entanglement properties of continuous variable states. We discuss the feasibility of our proposed scheme considering realistic experimental conditions such as on-off photodetectors with nonideal efficiency and imperfect single-photon sources.
\end{abstract}

\maketitle

\section{Introduction}
The superposition principle in quantum mechanics plays a crucial role in manifesting physical effects that go beyond the reach of classical descriptions.
A coherent superposition, when realized at the level of quantum operation rather than quantum state, can provide a useful tool for a number of applications, e.g., in quantum information processing. In the regime of continuous-variables (CVs), the superposition operation $\hat{a}\hat{a}^{\dag}\pm \hat{a}^{\dag}\hat{a}$ was proposed to test the commutation relation $[\hat{a},\hat{a}^{\dag}]=1$ \cite{Kim}, which was experimentally realized in a single-photon interferometric setting using thermal lights \cite{Zavatta}. ($\hat{a}^{\dag},\hat{a}$: bosonic creation and annihilation operators, respectively). This idea was extended to a proposal of implementing an arbitrary polynomial of photon-number
operators using multiple photon subtractions and additions \cite{Fiurasek0}. Furthermore, the superposition operation at a more elementary level of first-order field operators, i.e., $t\hat{a}+r\hat{a}^{\dag}$, was studied with possible applications to quantum-state engineering \cite{Lee} and entanglement (nonlocality) concentration \cite{Lee1}.
In addition to these {\it local} coherent operations, some {\it nonlocal} coherent operations were also investigated such as $\hat{a}+\hat{b}$ \cite{Grangier}, $\hat{a}^2+\hat{b}^2$ \cite{Kok}, and $\hat{a}^{\dag}+\hat{b}^{\dag}$ \cite{Fiurasek}.

In general, the working principle of realizing a coherent superposition operation is to erase the which-path information relevant to the operation under study.
For example, to implement the superposition operation $t\hat{a}+r\hat{a}^{\dag}$, one may place a beam splitter before detecting a photon in order to erase the information on whether the photon emerges from the path of photon subtraction $\hat{a}$ or photon addition $\hat{a}^{\dag}$ \cite{Lee}.
In this paper, we propose a second-order nonlocal operation  $\hat{a}^{\dag2}+e^{i\phi}\hat{b}^{\dag2}$, and its variants, via the effect of Hong-Ou-Mandel (HOM) interferometer \cite{Hong}. The HOM interference arises when two indistinguishable photons are each injected into the input modes of a 50:50 beam splitter: the output state from the beam splitter shows a bunching effect such that both photons appear together at one of two output modes, which may be attributed to the bosonic nature of photons.
The HOM interference can thus be used for the creation of a NOON state for $N=2$ deterministically, or the demonstration of purity of a single photon \cite{Santori, Beugnon}.
Conversely, with the beam splitting being a reversible unitary operation, if a single photon is detected for each output mode, the input state is inferred to be $\frac{1}{\sqrt{2}}\left(|2,0\rangle-|0,2\rangle\right)$.
When this HOM effect is employed in conjunction with two non-degenerate parametric amplifiers (NDPAs),
the operation $\hat{a}^{\dag2}-\hat{b}^{\dag2}$ can be implemented on two signal modes by projecting the idler modes to $\frac{1}{\sqrt{2}}\left(|2,0\rangle-|0,2\rangle\right)=\frac{1}{2}\left(\hat{c}^{\dag2}-\hat{d}^{\dag2}\right)|0,0\rangle$, as will be shown later.

The superposition operation  $\hat{a}^{\dag2}+e^{i\phi}\hat{b}^{\dag2}$ can be useful for a number of applications, and in this paper, we particularly discuss the generation of a NOON state with high particle number and the enhancement of entanglement properties. A NOON state is known to be a valuable resource for quantum lithography and quantum metrology, e.g., beating the shot-noise limit in an optical phase measurement \cite{Dowling}, and also for linear-optical quantum computing \cite{Knill}.  In Ref. \cite{Fiurasek}, the multiple use of the operation $\hat{a}^{\dag}+e^{i\phi}\hat{b}^{\dag}$ was proposed to produce a N00N state, which is, however, implemented on the condition of the non-detection events. Usually, the conditioning on non-detection may significantly suffer from nonideal detector efficiency, as one cannot know whether the nondetection is attributed to no photons being present or to a failure of the detector \cite{Lee3}. To overcome this difficulty, a heralded scheme implementing the superposition operation $\hat{a}^2+\hat{b}^2$ was also proposed, which, however, requires a Fock-state input $|N,N\rangle$ of large $N$ that is thereby rather demanding \cite{Kok,Hwang}.
On the other hand, our scheme makes use of the vacuum-state inputs $|0,0\rangle$, for which the superposition operation $\hat{a}^{\dag2}+e^{i\phi}\hat{b}^{\dag2}$ is consecutively applied. Alternatively, one may first prepare a $|2,0\rangle-|0,2\rangle$ state by injecting single photons into a 50:50 beam splitter and then applying $\hat{a}^{\dag}+e^{i\phi}\hat{b}^{\dag}$ or $\hat{a}^{\dag2}+e^{i\phi}\hat{b}^{\dag2}$ in a heralded fashion.
The latter alternative scheme is here investigated in some detail by including experimental imperfections such as on-off photodetectors with nonideal efficiency and imperfect single-photon sources.

This paper is organized as follows. In Sec. II, we propose how the second-order local and nonlocal superposition operations are implemented via the Hong-Ou-Mandel interference. In Sec. III, we investigate their applications for the generation of a NOON state with high particle number and the enhancement of entanglement properties.
In Sec. IV, we study in more detail the experimental feasibility of generating a $4004$ state in terms of the fidelity and the phase-sensitivity of a phase-measurement, and in Sec. V, our main results are summarized.

\section{Second-order coherent superposition operation via HOM interference}
In this section, we propose an optical method to implement a coherent superposition operation $\hat{a}^{\dag 2}+e^{i\phi}\hat{b}^{\dag 2}$ in a heralded fashion via the HOM interference. Our scheme is depicted in Fig. 1, where the success of the operation is heralded by the detection of a single photon at both photodetectors $\rm SPD_1$ and $\rm SPD_2$.

When a two-mode Fock-state input $|n\rangle_c|m\rangle_d$ is injected into a 50:50 beam splitter, the output state is given by
\begin{eqnarray}
\hat{B}_{cd}|n\rangle_c|m\rangle_d=\frac{(\hat{c}^{\dag}+\hat{d}^{\dag})^n(-\hat{c}^{\dag}+\hat{d}^{\dag})^m|0\rangle_{cd}}{\sqrt{2^{n+m}n!m!}}.
\end{eqnarray}
For the input $\{n,m\}=\{1,0\}$ or $\{0,1\}$, the output is the single photon entangled state $\frac{1}{\sqrt{2}}(|1,0\rangle\pm|0,1\rangle$.
On the other hand, for the input $\{n,m\}=\{1,1\}$, the output is a NOON state $\frac{1}{\sqrt{2}}(|2,0\rangle-|0,2\rangle)$. Conversely, as the beam splitting is a reversible unitary operation, these results imply that the detection of a single photon at $\rm SPD_1$ and no photons at $\rm SPD_2$, or vice versa, projects the input state to $\frac{1}{\sqrt{2}}(|1,0\rangle\pm|0,1\rangle)=\frac{1}{\sqrt{2}}(\hat{c}^{\dag}\pm\hat{d}^{\dag})|0,0\rangle$. On the other hand, the detection of a single photon each at both detectors $\rm SPD_1$ and $\rm SPD_2$ projects the input state to $\frac{1}{\sqrt{2}}(|2,0\rangle-|0,2\rangle)=\frac{1}{2}(\hat{c}^{\dag2}-\hat{d}^{\dag2})|0,0\rangle$.

When the above two-mode projective measurement is made on the idler modes from two independent NDPAs, the superposition operations $\hat{a}^{\dag}\pm\hat{b}^{\dag}$ and $\hat{a}^{\dag2}-\hat{b}^{\dag2}$ can be implemented on the signal modes.
This is because the interaction between the signal mode $a (b)$ and the idler mode $c (d)$ within the NDPA creates and annihilates photons in a pair-wise fashion. For example, the detection of two photons at the idler mode immediately implies the two-photon creation at the signal mode. Therefore, the projection on the idler modes is identically mapped to the projection on the signal modes. It is in a sense an entanglement swapping by a projective measurement \cite{Zukowski,Pan}: The two signal modes initially uncorrelated become entangled by the projection of the idler modes to an entangled state.

More rigorously, when an arbitrary two-mode state $|\psi\rangle_{ab}$ is injected into the signal modes of two NDPAs with both the idler modes in a vacuum state, the output state is given by
\begin{eqnarray}
&&\hat{S}_{ac}(\xi_1)\hat{S}_{bd}(\xi_2)|\psi\rangle_{ab}|0\rangle_{cd},\nonumber\\
&&=\exp(-\xi_1\hat{a}^{\dag}\hat{c}^{\dag}+\xi^*_1\hat{a}\hat{c})\exp(-\xi_2\hat{b}^{\dag}\hat{d}^{\dag}+\xi^*_2\hat{b}\hat{d})|\psi\rangle_{ab}|0\rangle_{cd}. \nonumber\\
\end{eqnarray}
Next, the 50:50 beam splitter with the transformations $\hat{c}\rightarrow \frac{1}{\sqrt{2}}(\hat{c}+\hat{d})$ and $\hat{d}\rightarrow \frac{1}{\sqrt{2}}(-\hat{c}+\hat{d})$ yields
\begin{eqnarray}
&&\hat{B}_{cd}\hat{S}_{ac}(\xi_1)\hat{S}_{bd}(\xi_2)|\psi\rangle_{ab}|0\rangle_{cd},\nonumber\\
&&=e^{-\frac{\hat{a}^{\dag}}{\sqrt{2}}(\hat{c}^{\dag}+\hat{d}^{\dag})e^{i\phi_1}\tanh{s_1}}
e^{-\frac{\hat{b}^{\dag}}{\sqrt{2}}(-\hat{c}^{\dag}+\hat{d}^{\dag})e^{i\phi_2}\tanh{s_2}}\nonumber\\
&&e^{-\hat{a}\hat{a}^{\dag}\ln(\cosh{s_1})}e^{-\hat{b}\hat{b}^{\dag}\ln(\cosh{s_2})}|\psi\rangle_{ab}|0\rangle_{cd},
\end{eqnarray}
where $\xi_1=s_1e^{\phi_1}$ and $\xi_2=s_2e^{\phi_2}$.
With the coincident detection of single photons at SPD1 and SPD2, the state is projected to
\begin{eqnarray}
|\Psi\rangle_{ab}&=&_{cd}\langle 11|\hat{B}_{cd}\hat{S}_{ac}(\xi_1)\hat{S}_{bd}(\xi_2)|\psi\rangle_{ab}|0\rangle_{cd} \nonumber\\
&=& \frac{1}{2}(\hat{a}^{\dag 2}e^{2i\phi_1}\tanh^2{s_1}-\hat{b}^{\dag 2}e^{2i\phi_2}\tanh^2{s_2})\nonumber\\
&&e^{-\hat{a}\hat{a}^{\dag}\ln(\cosh{s_1})}e^{-\hat{b}\hat{b}^{\dag}\ln(\cosh{s_2})}|\psi\rangle_{ab}.
\end{eqnarray}
Under the weak-coupling condition $s_1\approx s_2 \ll1$, the output field
is approximated to $|\Psi\rangle_{ab}\approx (\hat{a}^{\dag 2}+\hat{b}^{\dag 2}e^{i\phi})|\psi\rangle_{ab}$,
where $\phi=2(\phi_2-\phi_1)+\pi$. The phase $\phi$ can be controlled by adjusting the phase of pumping fields to the NDPAs, which determine the coupling constants $\xi_1$ and $\xi_2$ proportional to the second-order susceptibility of the nonlinear medium \cite{Walls}.

Obviously, the interferometric effect described above can be realized only if the optical paths from the two idler modes to the beam splitter have the same length in Fig. 1. In a pulsed-mode implementation of our scheme, the two down converters may be pumped with external fields repeatedly at the same predetermined times.
Then, the photo detections at SPD1 and SPD2 do not reveal the information on the origin of the photons, thus realizing the coherent operation $\hat{a}^{\dag 2}+\hat{b}^{\dag 2}e^{i\phi}$ on the input two-mode signal.

\begin{figure}
\centerline{\scalebox{0.4}{\includegraphics[angle=270]{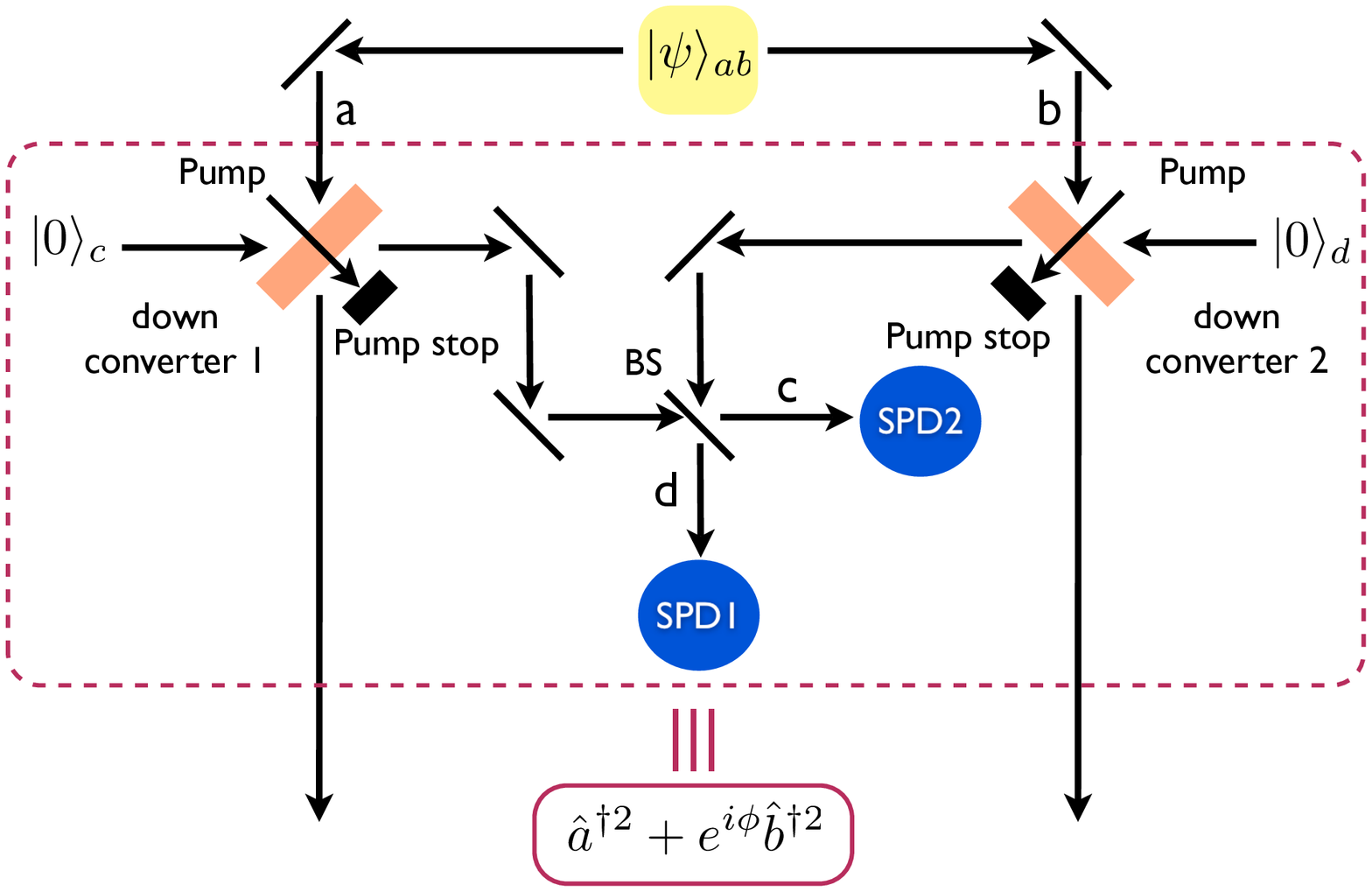}}}
\vspace{-1.2in}
\caption{Optical scheme to implement a coherent superposition operation $\hat{a}^{\dag 2}+e^{i\phi}\hat{b}^{\dag 2}$ on an arbitrary two-mode state $|\psi\rangle_{ab}$. BS is a 50:50 beam splitter; SPD1 and SPD2 are single photon detectors. The successful implementation of the superposition operation is heralded by the coincident detection of single photons at SPD1 and SPD2.}
\label{FIG_1}
\end{figure}

If desired, one can also change the ratio of the operations $\hat{a}^{\dag 2}$ and $\hat{b}^{\dag 2}$ in the superposition by inserting additional beam splitters between the down converters and the 50:50 beam splitter, as shown in Fig. 2 (a).
By the coincident detection of single photons at SPD1 and SPD2, together with the non-detection at the additional detectors PDs, the output state is now reduced to
\begin{eqnarray}
&&_{cd}\langle 11|_{ef}\langle 00|\hat{B}_{cd}\hat{B}_{ce}\hat{B}_{df}\hat{S}_{ac}(\xi_1)\hat{S}_{bd}(\xi_2)|\psi\rangle_{ab}|0\rangle_{cdef}\nonumber\\
&=& \frac{1}{2}(\hat{a}^{\dag 2}t^{*2}_1e^{2i\phi_1}\tanh^2{s_1}-\hat{b}^{\dag 2}t^{*2}_2e^{2i\phi_2}\tanh^2{s_2})\nonumber\\
&&e^{-\hat{a}\hat{a}^{\dag}\ln(\cosh{s_1})}e^{-\hat{b}\hat{b}^{\dag}\ln(\cosh{s_2})}|\psi\rangle_{ab},
\end{eqnarray}
where the beam splitter $\hat{B}_{ce}$ ($\hat{B}_{df}$) transforms the initial modes as $\hat{c}\rightarrow t_1\hat{c}+r_1\hat{e}$ and
$\hat{e}\rightarrow t_1\hat{e}-r_1\hat{c}$ ($\hat{d}\rightarrow t_2\hat{d}+r_2\hat{f}$ and
$\hat{f}\rightarrow t_2\hat{f}-r_2\hat{d}$ ). $t_{1(2)}$ and $r_{1(2)}$ denote the transmissivity and reflectivity of the beam splitters, respectively.
Under the weak-coupling condition $s_1\approx s_2 \ll1$, the output field is approximated to $(\hat{a}^{\dag 2}+e^{i\phi}\gamma\hat{b}^{\dag 2})|\psi\rangle_{ab}$, with the ratio given by $\gamma=\frac{t^{*2}_2}{t^{*2}_1}$.

The nonlocal superposition operation $\hat{a}^{\dag 2}+e^{i\phi}\gamma\hat{b}^{\dag 2}$ can also be transformed to a local superposition operation $\hat{a}^{2}+\gamma\hat{a}^{\dag 2}$, as shown in Fig. 2 (b), by replacing one down converter with a beam splitter of high transmissivity.
The output state now reads
\begin{eqnarray}
&&_{bd}\langle 11|_{ce}\langle 0|\hat{B}_{bd}\hat{B}_{bc}\hat{B}_{de}\hat{S}_{ad}\hat{B}_{ab}|\psi\rangle_a|0\rangle_{bcde}\nonumber\\
&&\approx -\frac{1}{2} [\hat{a}^2(\frac{r^*}{t})^2t^{*2}_1+\hat{a}^{\dag 2}\xi^2t^{* 2}_2]|\psi\rangle_a,
\end{eqnarray}
where $t$ ($r$) is the transmissivity (reflectivity) of $\hat{B}_{ab}$, and $t_{1(2)}$ is the transmissivity of $\hat{B}_{bc(de)}$.
At $(\frac{r^*}{t})^2 \approx \xi^2 \ll 1$, the output field is approximated to $(\hat{a}^{2}+\gamma\hat{a}^{\dag 2})|\psi\rangle_a$.
For the case of $\gamma=1$, one can simplify the implementation of the local operation $\hat{a}^{2}+\hat{a}^{\dag 2}$ by removing modes $c$ and $e$, the two beam splitters at the center, and the additional detectors PDs in Fig. 2 (b), which recovers a fully heralded scheme.

\begin{figure}
\centerline{\scalebox{0.4}{\includegraphics[angle=270]{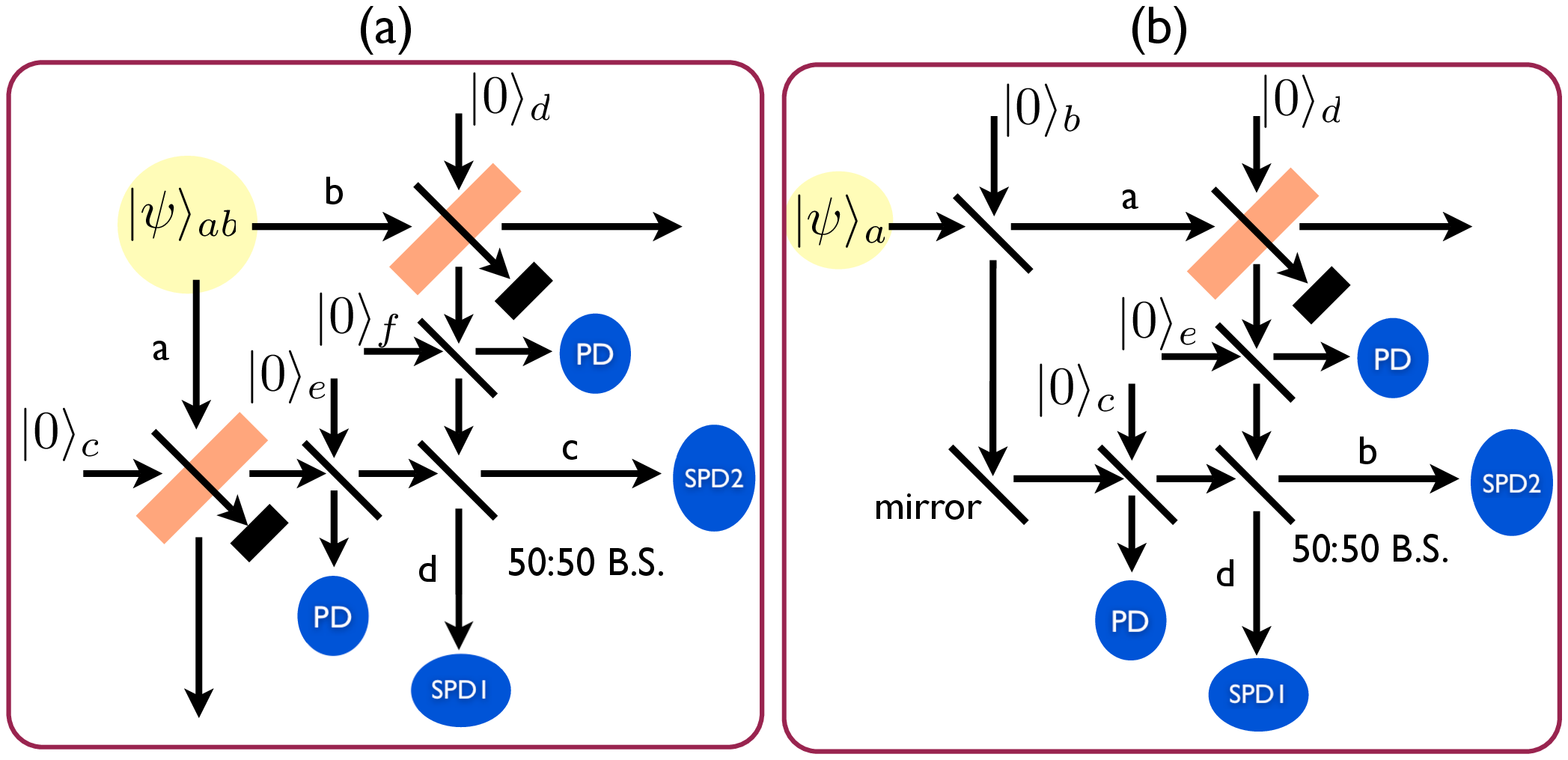}}}
\vspace{-1.5in}
\caption{
Optical schemes to implement (a) a generalized nonlocal operation $\hat{a}^{\dag 2}+e^{i\phi}\gamma\hat{b}^{\dag 2}$ with
$\gamma=\frac{t^{*2}_2}{t^{*2}_1}$ on an arbitrary two-mode state $|\psi\rangle_{ab}$ and (b) a local superposition operation $\hat{a}^2+\gamma\hat{a}^{\dag 2}$
on an arbitrary single-mode state $|\psi\rangle_a$. SPD1 and SPD2 are single-photon detectors, and PD indicates an on-off detector.
Under the condition of non-detection at the PDs, the coherent operations are successfully implemented
with the coincident single-photon detections at SPD1 and SPD2.}
\label{FIG_2}
\end{figure}

\section{Applications}
The second-order coherent superposition operations can be employed for a number of applications, and we particularly investigate the generation of NOON states and the enhancement of entanglement properties for CV states. First, as shown in Fig. 3, the coherent operation $\hat{a}^{\dag 2}+e^{i\phi}\hat{b}^{\dag 2}$, when consecutively applied, can produce a NOON state with high particle number for even $N$ as
\begin{eqnarray}
(\hat{a}^{\dag N}+\hat{b}^{\dag N})|0,0\rangle =\prod_{k=1}^{N/2}(\hat{a}^{\dag 2}+e^{i\phi_k}\hat{b}^{\dag 2})|0,0\rangle,
\end{eqnarray}
with the choice of $\phi_k=\frac{4\pi k}{N}$.
When an even-number NOON state is prepared, an odd-number NOON state can also be obtained by applying a coherent photon-subtraction $\hat{a}+\hat{b}$, that is, $(\hat{a}+\hat{b})\left(|N,0\rangle+|0,N\rangle\right)\sim|N-1,0\rangle+|0,N-1\rangle$.
We have previously seen in Ref. \cite{Lee3} that the high fidelity for an odd $N$ state can be achieved from the coherent photon-subtraction, even with a very low detector efficiency used for the heralding, if the initial even-$N$ NOON state can be generated with high fidelity. In the next section, we investigate in more detail the generation of NOON states considering experimental imperfections for a realistic application.
\begin{figure}
\centerline{\scalebox{0.35}{\includegraphics[angle=270]{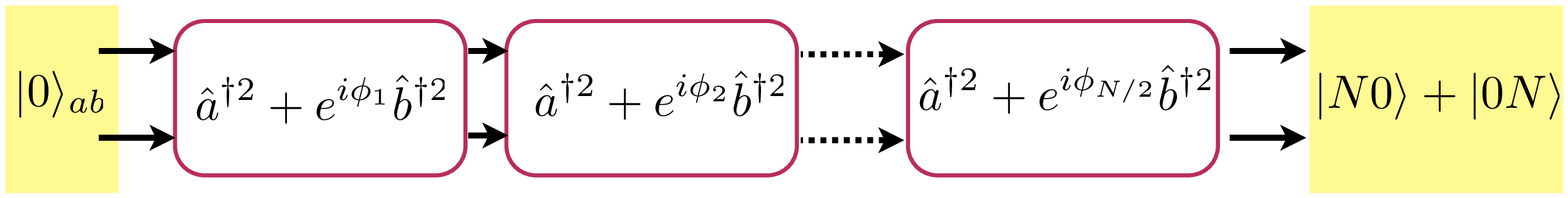}}}
\vspace{-2.3in}
\caption{Generation of NOON states via a successive application of the coherent operation $\hat{a}^{\dag 2}+e^{i\phi}\hat{b}^{\dag 2}$.}
\label{FIG_3}
\end{figure}

Second, the local coherent superposition operation $\hat{a}^2+\gamma\hat{a}^{\dag 2}$ can be useful to enhance the entanglement properties of
a CV entangled state, e.g., two-mode squeezed vacuum state,
$|S_{\rm TMSS}\rangle_{AB}=\sqrt{1-\lambda^2}\sum^{\infty}_{n=0}\lambda^n|n\rangle_A|n\rangle_B~(\lambda=\tanh{s})$.
In Fig. 4 (a), the degree of entanglement, which can be quantified by the von Neumann entropy of the reduced density operator for a pure state \cite{Bennett}, is compared between the states obtained with first-order and second-order superposition operations on the TMSS.
It is clearly seen that the entanglement is more improved by the second-order coherent operation than by the first-order coherent operation  $ t\hat{a}+r\hat{a}^{\dag}$ with $|t|^2+|r|^2=1$. Moreover, the second-order operation can significantly improve the Einstein-Podolsky-Rosen (EPR) correlation characterized by the condition $\Delta^2(\hat{x}_A-\hat{x}_B)+\Delta^2(\hat{p}_A+\hat{p}_B)<2$, where $\hat{x}_j=\frac{1}{\sqrt{2}}(\hat{a}_j+\hat{a}^{\dag}_j)$ and
$\hat{p}_j=\frac{1}{i\sqrt{2}}(\hat{a}_j-\hat{a}^{\dag}_j)$ ($j=A,B$) \cite{Duan}. In particular, the improvement of the EPR correlation is more enhanced by the second-order operation in the small-squeezing region, $0.08 < s < 0.47$, as shown in Fig. 4 (b), which may thus provide a practical advantage.

\begin{figure}
\centerline{\scalebox{0.36}{\includegraphics[angle=270]{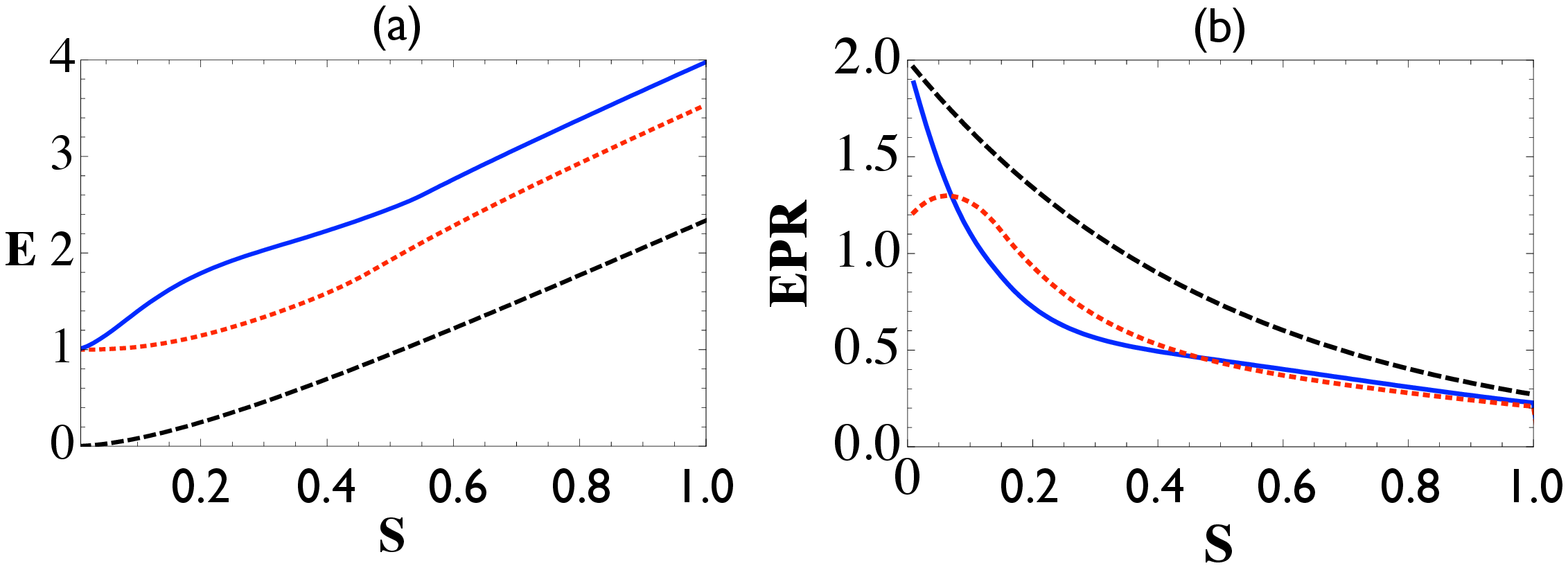}}}
\vspace{-1.7in}
\caption{(a) Entanglement quantified by the von Neumann entropy and (b) EPR correlation as functions of the squeezing parameter $s$ for the states:
$(t\hat{a}^2+r\hat{a}^{\dag 2})(t\hat{b}^2+r\hat{b}^{\dag 2})|S_{\rm TMSS}\rangle$ (blue solid line),
$(t\hat{a}+r\hat{a}^{\dag})(t\hat{b}+r\hat{b}^{\dag})|S_{\rm TMSS}\rangle$ (red dotted), and $|S_{\rm TMSS}\rangle$ (black dashed).
The value $r$ in each coherent operation ($t^2+r^2=1$) is optimized at each point of $s$.
}
\label{FIG_4}
\end{figure}

\section{Experimental feasibility}

\begin{figure}
\centerline{\scalebox{0.36}{\includegraphics[angle=270]{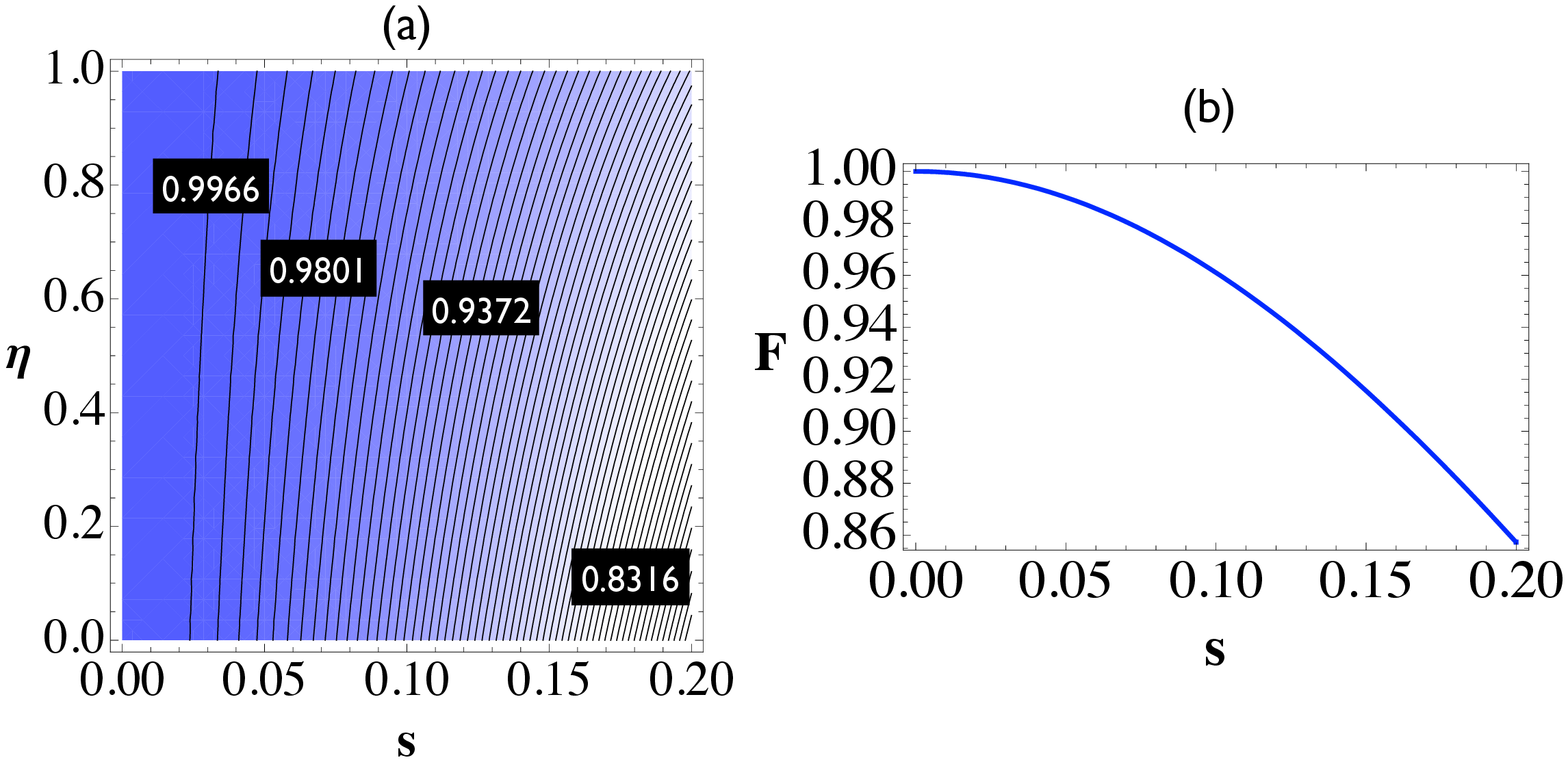}}}
\vspace{-1.2in}
\caption{Fidelity between the ideal NOON state $|4,0\rangle-|0,4\rangle$ and the state $\rho_{\rm out}$ [Eq. (8)] obtained by applying $\hat{a}^{\dag 2}+\hat{b}^{\dag 2}$, using on-off detectors with efficiency $\eta$, to a two-photon state $(|20\rangle-|02\rangle)/\sqrt{2}$ (a) as a function of the coupling strength $s$ of the NDPAs and the detector efficiency $\eta$ and (b) as a function of $s$ for $\eta=0.66$.
}
\label{FIG_5}
\end{figure}

In this section, we consider realistic experimental conditions in implementing our proposed scheme for the superposition operation $\hat{a}^{\dag 2}+e^{i\phi}\hat{b}^{\dag 2}$ and also the generation of a NOON state. Ideally, the second-order superposition operation is heralded by the detection of exactly one photon at both detectors SPD1 and SPD2 in Fig. 1. We here consider each SPD as an on-off detector with efficiency $\eta$ that can only distinguish two events, detection and non-detection, with no photon-number resolving. It can be characterized by a two-component positive-operator-valued-measure (POVM) \cite{Mog,Rossi},
$\hat{\Pi}_0=\sum_n (1-\eta)^n|n\rangle\langle n|$ (no click), and $\hat{\Pi}_1=\hat{I}-\hat{\Pi}_0$ (click).
Suppose that one first prepares a NOON state $|\psi_2\rangle\equiv\frac{1}{\sqrt{2}}\left(|2\rangle_a|0\rangle_b-|0\rangle_a|2\rangle_b\right)$ deterministically, to which the superposition operation $\hat{a}^{\dag 2}+e^{i\phi}\hat{b}^{\dag 2}$ heralded by nonideal on-off detectors is applied.
This yields an output state
\begin{eqnarray}
\rho_{\rm out}=\frac{{\rm Tr}_{cd}\{\hat{\Pi}_1^c\hat{\Pi}_1^d\hspace{0.2cm}{\hat U}_1\rho_{\rm in}{\hat U}_1^\dag\}}{{\rm Tr}_{abcd}\{\hat{\Pi}_1^c\hat{\Pi}_1^d\hspace{0.2cm}{\hat U}_1\rho_{\rm in}{\hat U}_1^\dag\}},
\end{eqnarray}
where $\rho_{\rm in}\equiv|\psi_2\rangle\langle\psi_2|_{ab}\otimes|0\rangle\langle0|_{cd}$ and ${\hat U}_1\equiv \hat{B}_{cd}\hat{S}_{ac}\hat{S}_{bd}$.
We evaluate the performance of our scheme by investigating the fidelity $F$ between $\rho_{\rm out}$ and the ideal NOON state $|\psi_4\rangle= \frac{1}{\sqrt{2}}\left(|4\rangle_a|0\rangle_b-|0\rangle_a|4\rangle_b\right)$, i.e. $F=\langle \psi_4|\rho_{\rm out}|\psi_4\rangle$.
In Fig. 5 (a), we plot the fidelity $F$ as a function of the coupling strength $s$ of the two NDPAs and the on-off detector efficiency $\eta$.
We see that a high fidelity is achievable even with a very low detector efficiency $\eta$, which is a usual practical advantage arising from a heralded scheme.
The fidelity decreases with the coupling strength $s$ because a larger $s$ makes multi-photon generations within the NDPAs more substantial and the on-off detectors cannot distinguish the multi-photons from an exact one-photon. Nevertheless, the fidelity remains considerably high to $s\sim0.2$ \cite{Alexei}.
In Fig. 5 (b), we plot the fidelity for the case of $\eta=0.66$, which is the detection efficiency currently available \cite{Achilles,Fitch,Brida}.
The fidelity is achieved at least above $0.86$ for the whole range of $s<0.2$.

In the above analysis, the initial state was assumed to be a pure two-photon state $|2\rangle_a|0\rangle_b-|0\rangle_a|2\rangle_b$, which can be generated by injecting a perfect single-photon state into each input-mode of a 50:50 beam-splitter. Now, let us address the case of imperfect single-photon sources, denoted by $\rho_s=(1-p)|0\rangle\langle 0|+p|1\rangle\langle 1|$, which are a mixture of a vacuum state and a single-photon state.
When this imperfect state $\rho_s$ is injected for each input of a 50:50 beam-splitter and the coherent operation $\hat{a}^{\dag 2}+e^{i\phi}\hat{b}^{\dag 2}$ is subsequently applied to the output from the beam splitter, we obtain a non-ideal NOON state $\rho_{\rm out,r}$ for $N=4$,
\begin{eqnarray}
\rho_{\rm out,r}=\frac{{\rm Tr}_{cd}\{\hat{\Pi}_1^c\hat{\Pi}_1^d\hspace{0.2cm}{\hat U}_2\rho_{\rm in,r}{\hat U}_2^\dag\}}{{\rm Tr}_{abcd}\{\hat{\Pi}_1^c\hat{\Pi}_1^d\hspace{0.2cm}{\hat U}_2\rho_{\rm in,r}{\hat U}_2^\dag\}},
\end{eqnarray}
where $\rho_{\rm in,r}\equiv\rho^a_{s}\otimes\rho^b_{s}\otimes|0\rangle\langle0|_{cd}$ and ${\hat U}_2\equiv \hat{B}_{cd}\hat{S}_{ac}\hat{S}_{bd}\hat{B}_{ab}$.
We compare $\rho_{\rm out,r}$ with the ideal state $|\psi_4\rangle= \frac{1}{\sqrt{2}}\left(|4\rangle_a|0\rangle_b-|0\rangle_a|4\rangle_b\right)$ in terms of fidelity. Furthermore, we also investigate the phase sensitivity $\Delta\varphi$ of a phase-measurement using the output state $\rho_{\rm out,r}$ that can be obtained via a Mach-Zehnder interferometer.
The interferometer can be designed to estimate the phase difference $\varphi$ between two optical paths by measuring the photon-number parity of an output field \cite{Dowling}. The sensitivity $\Delta\varphi$ in this case is given by $\Delta\varphi=\Delta\Pi_b/|\frac{\partial \langle \hat{\Pi}_b\rangle}{\partial \varphi}|$, where $\hat{\Pi}_b=(-1)^{\hat{b}^{\dag}\hat{b}}$ corresponds to the parity of an output field from the interferometer.
In Fig. 6 (a), we show the phase sensitivity as a function of single-photon source efficiency $p$ and on-off detector efficiency $\eta$ at the coupling strength $s=0.05$ of the NDPAs. The colored region represents the phase-sensitivity below the shot-noise limit using $\rho_{\rm out,r}$, for which the single-photon efficiency $p>0.68$ is required. We note that the value of $p=0.69$ was previously reported in the experiment of Ref. \cite{Lvovsky}.
In Fig. 6 (b), we also show the fidelity as a function of $p$ and $\eta$, which can reach the value of, e.g., 0.595 with $p=0.69$. Both the phase-sensitivity $\Delta \varphi$ and the fidelity $F$ are significantly affected by the single-photon efficiency $p$, but they are very insensitive to the efficiency $\eta$ of the on-off photodetectors.

\begin{figure}
\centerline{\scalebox{0.36}{\includegraphics[angle=270]{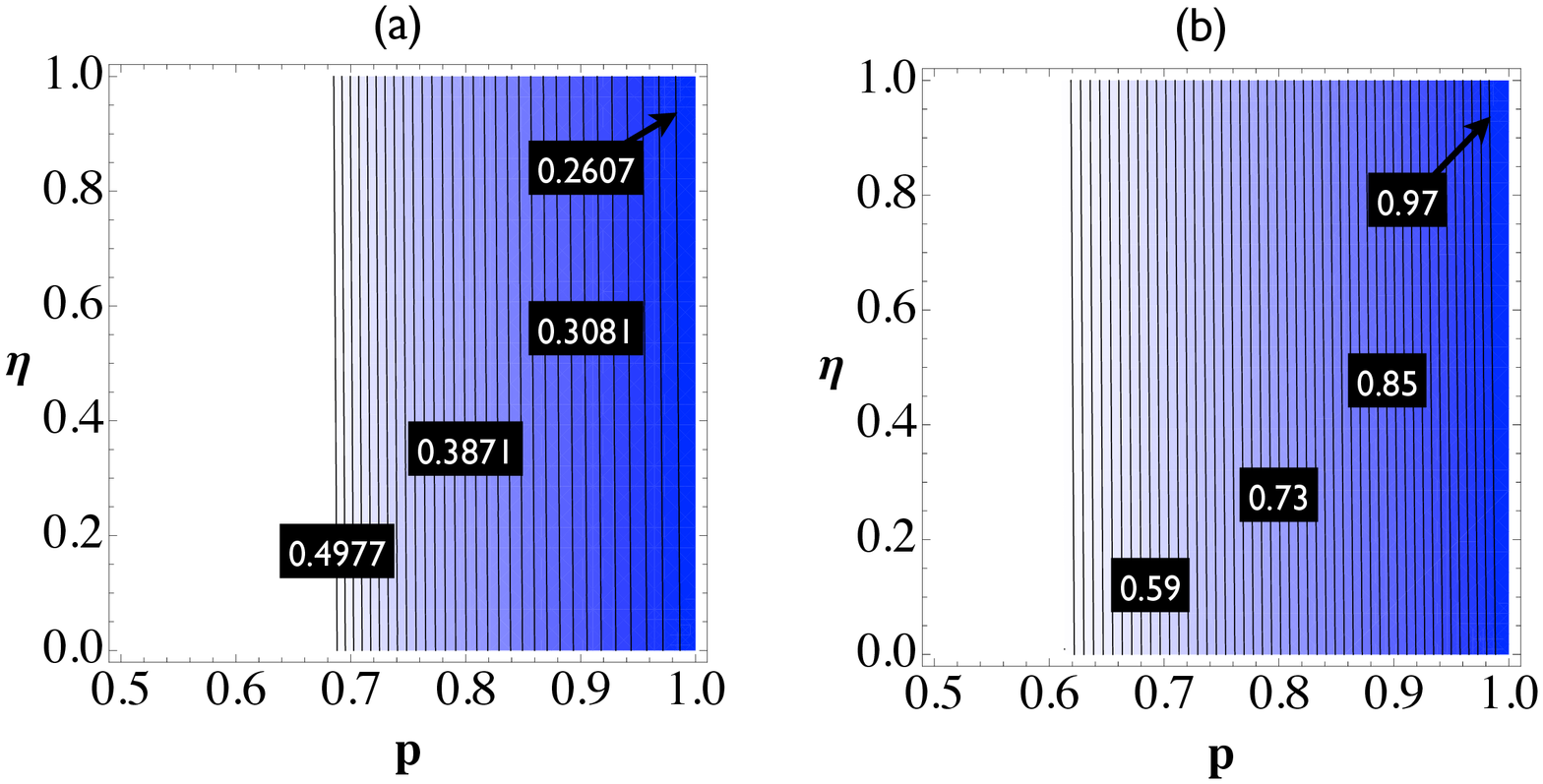}}}
\vspace{-1.2in}
\caption{(a) Phase sensitivity $\Delta \varphi$ of the phase measurement using the nonideal state $\rho_{\rm out,r}$ [Eq.(9)] and (b) fidelity $F$ between the ideal state $(|40\rangle-|04\rangle)/\sqrt{2}$ and $\rho_{\rm out,r}$ as functions of single-photon source efficiency $p$ and on-off detector efficiency $\eta$. (See main texts.) The coupling strength of the NDPAs is $s=0.05$. The colored region represents the condition for (a) $\Delta\varphi <0.5$ (below shot-noise level) and (b) $F\ge0.5$.
}
\label{FIG_6}
\end{figure}

\section{Summary}
We have proposed an optical scheme to implement a second-order nonlocal superposition operation $\hat{a}^{\dag 2}+e^{i\phi}\hat{b}^{\dag 2}$ and its variants in a heralded fashion via the HOM effect. We also investigated the application of these superposition operations to the generation of NOON states with high particle number and the entanglement concentration of a CV entangled state. Furthermore, we considered experimental imperfections such as on-off photodetectors with nonideal efficiency and imperfect single-photon sources to demonstrate the feasibility of our proposed scheme. In view of the experimental capacity reported, e.g., in Refs. \cite{Zavatta,Grangier,Alexei,Lvovsky}, our proposal can provide a potentially useful tool for various quantum information tasks.

\begin{acknowledgments}
S.Y.L. thanks J. Bae for a helpful discussion.
This work is supported by NPRP Grant 08-043-1-011 from the Qatar National Research Fund.
HN also acknowledges support from a research fellowship from the Alexander von Humboldt Foundation.
\end{acknowledgments}

\end{document}